# Precise Stock Price Prediction for Optimized Portfolio Design Using an LSTM Model

Jaydip Sen[1], Sidra Mehtab[2], Abhishek Dutta[3], Saikat Mondal[4]
Dept. of Data Science
Praxis Business School
Kolkata, India
emails: [1]jaydip.sen@acm.org, [2]smhetab@acm.org, [3]duttaabhishek0601@gmail.com, [4]sikatmo@gmail.com

*Abstract*— **Accurate prediction of future prices of stocks is a difficult task to perform. Even more challenging is to design an optimized portfolio of stocks with the identification of proper weights of allocation to achieve the optimized values of return and risk. We present optimized portfolios based on the seven sectors of the Indian economy. The past prices of the stocks are extracted from the web from January 1, 2016, to December 31, 2020. Optimum portfolios are designed on the selected seven sectors. An LSTM regression model is also designed for predicting future stock prices. Five months after the construction of the portfolios, i.e., on June 1, 2021, the actual and predicted returns and risks of each portfolio are computed. The predicted and the actual returns indicate the very high accuracy of the LSTM model.**

## I. INTRODUCTION

The design of optimized portfolios has remained a research topic of broad and intense interest among the researchers of quantitative and statistical finance for a long time. An optimum portfolio allocates the weights to a set of capital assets in a way that optimizes the return and risk of those assets. Markowitz in his seminal work proposed the mean-variance optimization approach which is based on the mean and covariance matrix of asset returns [1]. Despite the elegance in its theoretical framework, the mean-variance theory of portfolio has some major limitations. One of the major problems is the adverse effects of the estimation errors in its expected returns and covariance matrix on the performance of the portfolio. Since it is extremely challenging to accurately estimate the expected returns of an asset from its historical prices, it is a popular practice to use either a minimum variance portfolio or an optimized risk portfolio with the maximum Sharpe ratio as better proxies for the expected returns. However, due to the inherent complexity, several factors have been used to explain the expected returns.

This paper proposes an algorithmic method for designing efficient portfolios by selecting stocks from seven sectors of the National Stock Exchange (NSE) of India. Based on the report of the NSE on July 30, 2021, the five most significant stocks of each of the seven chosen sectors are first identified [2]. Portfolios are designed for the sectors optimizing the risks and returns. The past prices of these thirty stocks for the past five years are extracted using Python from the Yahoo Finance site. To aid the portfolio construction, an LSTM model is designed for predicting future stock prices and future returns of the portfolios for different forecast horizons. Five months after the portfolios are constructed, the actual returns and the predicted returns by the LSTM model are compared to evaluate the accuracy of the predictive model and to estimate the returns and risks associated with the seven sectors. The seven sectors studied in the work are auto, consumer durable, healthcare, information technology, metal, oil and gas, and FMCG.

The main contribution of the current work is threefold. First, it presents an approach to designing robust and optimum portfolios for seven sectors of India. The results of the portfolios may serve as a guide to investors in the stock market for making profitable investments in the stock market. Second, a precise deep learning-based regression model is proposed exploiting the power of LSTM architecture for predicting future stock prices for robust portfolio design. Third, the returns of the portfolios highlight the current profitability of investment and the volatilities of the seven sectors studied in this work.

The paper is structured as follows. In Section II, some existing works on portfolio design and stock price prediction are discussed briefly. Section III presents a description of the methodology followed in the work in a systematic manner. Section IV discusses the LSTM model. Section V presents the results of different portfolios and the predictions of future stock prices by the LSTM models. Section VI concludes the paper.

## II. RELATED WORK

Due to the challenging nature of the problems and their impact on real-world applications, several propositions exist in the literature for stock price prediction and robust portfolio design for optimizing returns and risk in a portfolio. The use of predictive models based on learning algorithms and deep neural net architectures for price stock price prediction is quite popular of late [3-6]. Hybrid models are also showcased that combine learning-based systems with the sentiments in the unstructured data on the web [7-9]. The use of multi-objective optimization and eigen portfolios using principal component analysis in portfolio design has also been proposed by some researchers [10-12]. The shortcomings of the optimum risk portfolio originally proposed by Markowitz have been addressed by introducing cardinality constraints [13-14]. Further, genetic algorithms, fuzzy logic, and swarm intelligence are some approaches to portfolio design.

In the current work, the min-variance approach is followed to build optimized portfolios for seven sectors. Using the past stock prices for five years from 2016 to 2020, seven portfolios are built. An LSTM model is then built for predicting the future prices of the stocks in each portfolio. Five months after the portfolio construction, the actual return for each portfolio and the return predicted by the LSTM model are computed. The results are analyzed for understanding the profitability of the sectors.

### III. Data and Methodology

It is stated in Section I that the objective of the current work is to design robust portfolios for seven critical sectors of the Indian economy. The second goal is to evaluate the accuracy of the LSTM model in predicting future stock prices and future returns and risks associated with each portfolio. The return-risk analysis also provides us with insights into the profitability and volatility of each sector and the investments in them. The Python programming language has been used in developing the proposed system. The Tensorflow and Keras frameworks are also used. This section presents the details of the six-step approach followed in the work in designing the proposed system. The steps are as follows.

#### A. Selection of the Stocks

Seven important sectors from the NSE of India are chosen first. The sectors are (i) auto, (ii) consumer durable, (iii) healthcare, (iv) information technology, (v) metal, (vi) oil and gas, and (vii) FMCG. Based on the criticality of stock in a particular sector, a weight is assigned to the stock which is used in deriving the aggregate sectoral index. The five most significant stocks for each sector are chosen based on the report published by the NSE on July 30, 2021 [2].

#### B. Data Acquisition

For each sector, the historical prices of the five most critical stocks are extracted using the DataReader function of the data sub-module of the pandas_datareader module in Python. The stock prices are extracted from the web from Jan 1, 2016, to Dec 31, 2020. There are five features in the stock data: open, high, low, close, volume, and adjusted_close. The current work is a univariate analysis, and hence, the variable close is chosen as the only variable of interest.

#### C. Computation of Return and Volatility

The percentage changes in the successive close values yield the daily returns for a stock. For computing the daily returns, the pct_change function of Python is used. Based on the returns on the daily basis, the daily and yearly volatilities of the five stocks of every sector are computed. Assuming that there are 250 operational days in a calendar year, the annual volatility values for the stocks are found by multiplying the daily volatilities by a square root of 250. The annual volatility indicates the risk associated with stocks from an investor's angle. The Python function std is used for computing the volatility.

#### D. Construction of the Minimum Risk Portfolios

At this step, for each sector, the minimum risk portfolio is designed. The portfolio with the minimum variance is referred to as the minimum variance portfolio. In order to identify the portfolio with the minimum variance for a given sector, first, the efficient frontier for many possible portfolios for that sector is plotted. The efficient frontier for a given sector represented the contour of a large number of portfolios on which the returns and the risks are plotted along the *y*-axis and the *x*-axis, respectively. The points on an efficient frontier have the property that they are the portfolios that yield the maximum return for a given risk, or they introduce the minimum risk for a given return. The left-most point on the efficient frontier depicts the point of minimum risk. For plotting the efficient frontier of a portfolio, weights are assigned randomly to the ten stocks over a loop which is iterated over 10,000 rounds in a Python program.

#### E. Identifying the Optimum Risk Portfolio

Minimum risk portfolios are rarely adopted in practice, and a risk-return optimization is done. For optimizing the risk, Sharpe Ratio (SR) is used, as derived from (1).

$$R = \frac{current\ portfolio\ return - risk\ free\ portfolio\ return}{current\ portfolio\ standard\ deviation} \quad (1)$$

In other words, the Sharpe Ratio optimizes the return and the risk by yielding a substantially higher return with a very marginal increase in the risk. The portfolio with a risk of 1% is assumed to be risk-free.

#### F. Computing the Actual and Predicted Returns

Using the training dataset from January 1, 2016, to December 31, 2020, two portfolios for each sector are built– a minimum risk portfolio and an optimal risk portfolio. On January 1, 2021, a fictitious investor is created who invests a capital of Indian Rupees (INR) of 100000 for each sector based on the recommendation of the optimal portfolio structure for the corresponding sector. Note that the amount of INR 100000 is just for illustrative purposes only. The analysis will not be affected either by the currency or by the amount. To compute the future values of the stock prices and hence to predict the future value of the portfolio, a regression model is built using the LSTM deep learning architecture. On May 31, 2021, using the LSTM model, the stock prices for June 1, 2021, are predicted (i.e., a forecast horizon of one day is used). Based on the predicted stock values, the predicted rate of return for each portfolio is determined. Finally, on June 1, 2021, when the actual prices of the stocks are known, the actual rates of return are computed. The predicted and actual rates of return for the portfolios are compared to evaluate the profitability of the portfolios and the prediction accuracy of the LSTM model.

## IV. THE MODEL DESIGN

As explained in Section III, the stock prices are predicted with a forecast horizon of one day, using an LSTM model. This section presents the details of the architecture and the choice of various parameters in the design. In the following, a very brief discussion on the fundamentals of LSTM networks and the effectiveness of these networks in interpreting sequential data are discussed and the details of the model are presented.

LSTM is an extended and advanced, recurrent neural network (RNN) with a high capability of interpreting and predicting future values of sequential data like time series of stock prices or text [15]. LSTM networks are able to maintain their state information in some specially designed memory cells or gates. The networks carry out an aggregation operation on the historical state stored in the forget gates with the current state information to compute the future state information. The information available at the current time slot is received at the input gates. Using the results of aggregation at the forget and the input gates, the network yields the next round's predicted result. The predicted value is available at the output gates [15].

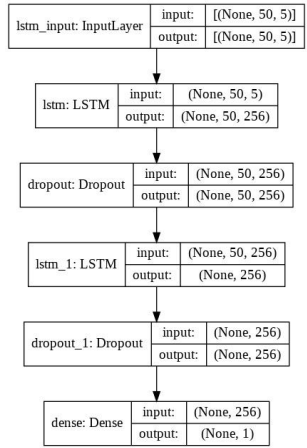

Figure 1. The schematic diagram of the LSTM model.

An LSTM model is designed and fine-tuned for predicting future stock prices. The schematic design of the model is exhibited in Fig. 1. The model uses daily close prices of the stock of the past 50 days as the input. The input data of 50 days with a single feature (i.e., close values) is represented by the data shape of (50, 1). The input layer forwards the data to the first LSTM layer. The LSTM layer is composed of 256 nodes. The output from the LSTM layer has a shape of (50, 256). Thus, each node of the LSTM layer extracts 256 features from every record in the input data. A dropout layer is used after the first LSTM layer that randomly switches off the output of thirty percent of the nodes in the LSTM to avoid model overfitting. Another LSTM layer with the same architecture as the previous one receives the output from the first and applies a dropout rate of thirty percent. A dense layer consisting of 256 nodes receives the output from the second LSTM layer. The dense layer has a single node at its output that produces the predicted value of the close price. The forecast horizon can be adjusted to different values by adjusting a tunable parameter. A forecast horizon of one day is used so that a prediction for the next day is made. To train and validate the model, a batch size of 64 and 100 epochs is used. Except for the output layer, the *rectified linear unit* (ReLU) activation function is used for all layers. At the final layer that produces the output, the sigmoid activation function is used. The loss and the accuracy during training and validation are measured using the Huber loss function and the *mean absolute error* (MAE) function, respectively. The hyperparameter values used in the network are all determined using the grid search method [15].

## V. RESULTS

In this section, the results of the portfolios of the seven sectors are presented and analyzed in detail. The chosen sectors for study are (i) *auto*, (ii) *consumer durable*, (iii) *healthcare*, (iv) *information technology* (IT), (v) *metal*, (vi) *oil and gas*, and (vii) FMCG. The model training and validation are carried out on the Google Colab platform.

### A. Auto Sector

The five significant stocks of this sector and their respective weights used in the sectoral index computation are Maruti Suzuki (MSU): 18.89, Mahindra and Mahindra (MMH): 15.51, Tata Motors (TMO): 11.46, and Hero MotoCorp (HMC): 7.83 [2]. Table I exhibits the wights allocated by the two portfolio strategies, and their return and risk values computed on Jun 1, 2021.

TABLE I   AUTO SECTOR PORTFOLIOS

| Stocks | Min Risk | Opt Risk |
|---|---|---|
| MSU | 0.1769 | 0.7364 |
| MMH | 0.2204 | 0.1126 |
| TMO | 0.0234 | 0.0184 |
| BAJ | 0.4839 | 0.1304 |
| HMC | 0.0955 | 0.0022 |
| Portfolio Annual Return | 8.69 % | 13.26 % |
| Portfolio Annual Risk | 23.38 % | 27.57 % |

TABLE II   ACTUAL AND PREDICTED RETURN OF AUTO PORTFOLIO

| Stock | Date: Jan 1, 2021 | | | Date: Jun 1, 2021 | | | |
|---|---|---|---|---|---|---|---|
|  | *Amt Invstd* | *Act Price* | *No of Stock* | *Act Price* | *Act Val* | *Pred Price* | *Pred Val* |
| MSZ | 73640 | 7691 | 9.57 | 7091 | 667861 | 7117 | 68110 |
| MMH | 11260 | 732 | 15.38 | 806 | 12396 | 812 | 12489 |
| TMO | 1840 | 187 | 9.84 | 318 | 3129 | 319 | 3139 |
| BAJ | 13040 | 3481 | 3.75 | 4239 | 15896 | 4177 | 15664 |
| HMC | 220 | 3103 | 0.07 | 2977 | 208 | 3022 | 212 |
| Total | 100000 | | | | 99490 | | 99614 |
| ROI | Actual: -0.51%   Predicted: -0.37% | | | | | | |

Table II presents the actual and the predicted return of the optimum portfolio over five months (i.e., from January 1, 2021, to June 1, 2021) as computed on June 1, 2021. Fig. 2 shows the efficient frontier, the min risk portfolio, and the opt. portfolio of the auto sector. As an illustration, Fig 3 depicts the actual prices and the corresponding predicted prices of the leading stock in this sector, i.e., MSU, from Jan 1, 2021, to Jun 1, 2021.

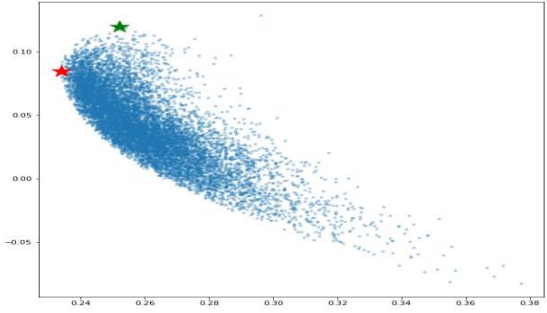

Figure 2. The min risk (red star) and the opt risk (green star) portfolios of the auto sector built on Jan 1, 2021. The risk and return are depicted on the x-and the y-axis, respectively.

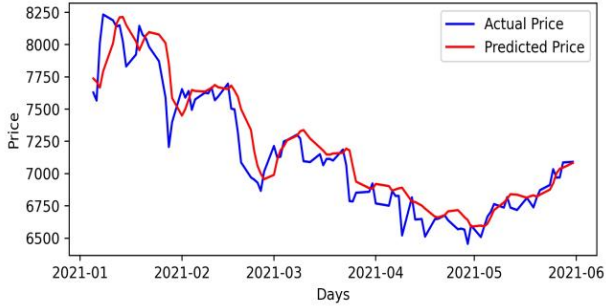

Figure 3. The act vs the pred values of the Maruti Suzuki (MSZ) stock as predicted by the LSTM model for the period: Jan 1 – Jun 1, 2021.

TABLE IV  CONS. DUR. SECTOR PORTFOLIOS

| Stocks | Min Risk | Opt Risk |
|---|---|---|
| TIT | 0.2094 | 0.3468 |
| HVL | 0.2375 | 0.0224 |
| VLT | 0.1665 | 0.0257 |
| CRP | 0.2552 | 0.1116 |
| DIX | 0.1312 | 0.4934 |
| Portfolio Annual Return | 45.91 % | 72.55% |
| Portfolio Annual Risk | 22.12 % | 27.45 % |

### B. Consumer Durable Sector

The five significant stocks and their corresponding weights used in computing the sectoral index are of this sector are Titan Company (TIT): 31.53, Havells India (HVL): 12.22, Voltas (VLT): 11.04, Crompton Greaves Consumer Electricals (CRP): 9.82, and Dixon Technologies India (DIX): 6.81 [2]. Tables IV and V show the results of the consumer durable portfolio. Fig 4 depicts the actual and predicted prices of the leading stock, TIT.

TABLE V  ACT AND PRED RETURNS OF CONS. DUR. PORTFOLIO

| Stock | Date: Jan 1, 2021 | | | Date: Jun 1, 2021 | | | |
|---|---|---|---|---|---|---|---|
| | Amt Invstd | Act Price | No of Stock | Act Price | Act Val | Pred Price | Pred Val |
| TIT | 34682 | 1559 | 22.25 | 1591 | 35400 | 1583 | 35222 |
| HVL | 2246 | 910 | 2.47 | 1029 | 2542 | 1033 | 2552 |
| VLT | 2571 | 831 | 3.09 | 1013 | 3130 | 1011 | 3124 |
| CRP | 11160 | 378 | 29.52 | 398 | 11749 | 394 | 11631 |
| DIX | 49341 | 2724 | 18.11 | 4107 | 74378 | 4101 | 74269 |
| Total | 100000 | | | | 127199 | | 126798 |
| Return | Actual: 27.20 %  Predicted: 26.80 % | | | | | | |

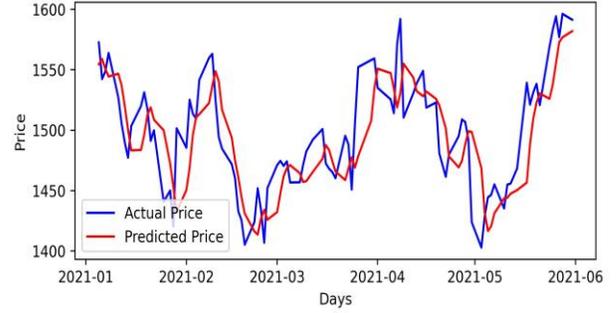

Figure 4. The act vs the pred values of the Titan Company (TIT) stock as predicted by the LSTM model for the period: Jan 1 – Jun 1, 2021.

TABLE VI  HEALTHCARE SECTOR PORTFOLIOS

| Stocks | Min Risk | Opt Risk |
|---|---|---|
| SNP | 0.1128 | 0.0055 |
| DRL | 0.2633 | 0.1470 |
| DVL | 0.1028 | 0.5727 |
| CPL | 0.2902 | 0.0233 |
| APL | 0.2309 | 0.2516 |
| Portfolio Annual Return | 19.75% | 38.15 % |
| Portfolio Annual Risk | 21.06 % | 26.41 % |

### C. Healthcare Sector

The five significant stocks and their weights in the healthcare sector are as follows. Sun Pharmaceuticals Industries (SNP): 14.90, Dr. Reddy's Lab (DRL): 13.31, Divi's Lab (DVL): 11.04, Cipla (CPL): 9.96, and Apollo Hospitals Enterprise (APL): 6.59 [2]. Tables VI and VII present the performance of the portfolio of this sector. Fig. 5 exhibits the actual prices vs their corresponding predicted prices of the SNP stock, the leading stock of the healthcare sector.

TABLE VII  ACT AND PRED RETURNS OF HEALTHCARE PORTFOLIO

| Stock | Date: Jan 1, 2021 | | | Date: Jun 1, 2021 | | | |
|---|---|---|---|---|---|---|---|
| | Amt Invstd | Act Price | No of Stock | Act Price | Act Val | Pred Price | Pred Val |
| SNP | 548 | 596 | 0.92 | 671 | 617 | 669 | 615 |
| DRL | 14698 | 5241 | 2.80 | 5317 | 14888 | 5295 | 14826 |
| DVL | 57268 | 3849 | 14.88 | 4220 | 62794 | 4232 | 62972 |
| CPL | 2328 | 827 | 2.81 | 946 | 2658 | 956 | 2686 |
| APL | 25158 | 2415 | 10.42 | 3240 | 33761 | 3286 | 34240 |
| Total | 100000 | | | | 114718 | | 115339 |
| Return | Actual: 14.72 %  Predicted: 15.34 % | | | | | | |

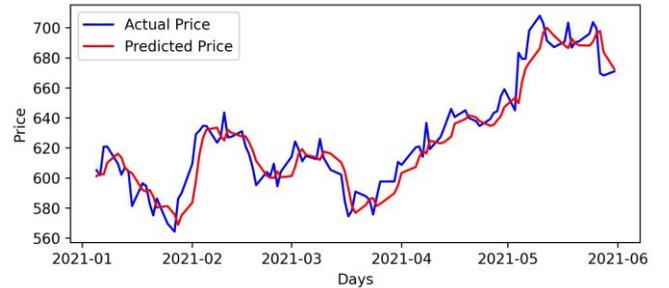

Figure 5. The act vs the pred values of the Sun Pharmaceutical (SNP) stock predicted by the LSTM model for the period: Jan 1– Jun 1, 2021.

TABLE VIII  IT Sector Portfolios

| Stocks | Min Risk | Opt Risk |
|---|---|---|
| IFY | 0.1452 | 0.2719 |
| TCS | 0.2385 | 0.2705 |
| WIP | 0.3496 | 0.2693 |
| TEM | 0.0908 | 0.0021 |
| HCL | 0.1758 | 0.1861 |
| Portfolio Annual Return | 24.52 % | 25.49 % |
| Portfolio Annual Risk | 20.82 % | 21.18 % |

TABLE IX  Act and Pred Returns of IT Sector Portfolio

| Stock | Date: Jan 1, 2021 | | | Date: Jun 1, 2021 | | | |
|---|---|---|---|---|---|---|---|
| | Amt Invstd | Act Price | No of Stock | Act Price | Act Val | Pred Price | Pred Val |
| IFY | 27192 | 1260 | 21.58 | 1387 | 29931 | 1413 | 30493 |
| TCS | 27052 | 2928 | 9.24 | 3153 | 29134 | 3151 | 29115 |
| WIP | 26930 | 388 | 69.41 | 543 | 37690 | 549 | 38106 |
| TEM | 214 | 978 | 0.22 | 1031 | 227 | 1029 | 226 |
| HCL | 18612 | 951 | 19.57 | 951 | 18611 | 962 | 18826 |
| Total | 100000 | | | | 115593 | | 116766 |
| ROI | Actual: 15.59 %  Predicted: 16.77 % | | | | | | |

### D. Information Technology (IT) Sector

The five important stocks and their corresponding weights used in deriving the overall sectoral index are Infosys (IFY): 25.10, Tata Consultancy Services (TCS): 24.76, Wipro (WIP): 12.40, Tech Mahindra (TEM): 9.69, and HCL Technologies (HCL): 9.08 [2]. Tables VIII and IX exhibit the results of this sector's portfolios. Fig. 6 shows the actual and the predicted prices of the leading stock, IFY.

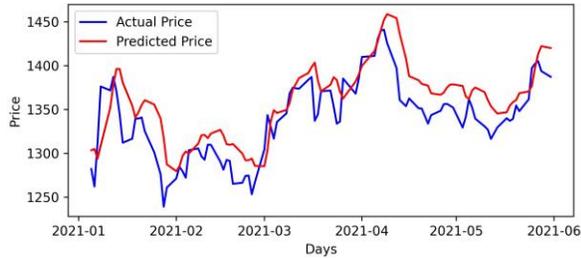

Figure 6. The act vs the pred values of the Infosys (IFY) stock predicted by the LSTM model for the period: Jan 1– Jun 1, 2021.

TABLE XIV  Metal Sector Portfolios

| Stocks | Min Risk | Opt Risk |
|---|---|---|
| TSL | 0.2206 | 0.0922 |
| JSW | 0.4249 | 0.1675 |
| HIN | 0.1430 | 0.0389 |
| ADE | 0.1537 | 0.6935 |
| VDN | 0.0578 | 0.0078 |
| Portfolio Annual Return | 32.58 % | 68.79 % |
| Portfolio Annual Risk | 32.54 % | 41.05 % |

TABLE XV  Act and Pred Returns of Metal Portfolio

| Stock | Date: Jan 1, 2021 | | | Date: Jun 1, 2021 | | | |
|---|---|---|---|---|---|---|---|
| | Amt Invstd | Act Price | No of Stock | Act Price | Act Val | Pred Price | Pred Val |
| TSL | 9220 | 643 | 14.34 | 1101 | 15788 | 1080 | 15487 |
| JSW | 16750 | 390 | 42.95 | 695 | 29850 | 674 | 28948 |
| HIN | 3890 | 238 | 16.34 | 395 | 6454 | 378 | 6177 |
| ADE | 69350 | 491 | 141.24 | 1416 | 199996 | 1341 | 189403 |
| VDN | 790 | 160 | 4.94 | 268 | 1324 | 273 | 1349 |
| Total | 100000 | | | | 253412 | | 241364 |
| Return | Actual: 153.41 %  Predicted: 141.36 % | | | | | | |

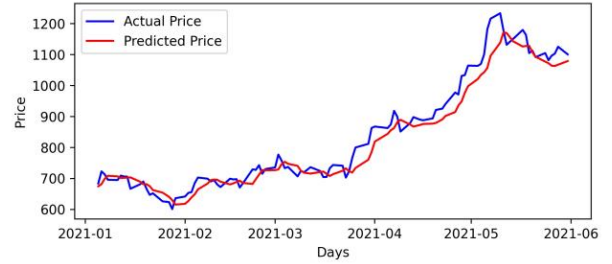

Figure 7. The act vs the pred values of the Tata Steel (TSL) stock predicted by the LSTM model for the period: Jan 1– Jun 1, 2021.

### E. Metal Sector

The five significant stocks in this sector with their respective contributions to the sectoral index are Tata Steel (TSL): 22.02, JSW Steel (JSW): 17.29, Hindalco Industries (HIN): 14.48, Adani Enterprises (ADE): 9.10, and Vedanta (VDN): 8.72 [2]. Tables XIV and XV present the results of the portfolios of the metal sector. Fig. 7 shows the actual and the predicted prices of the leading stock, TSL.

TABLE XVI  Oil & Gas Sector Portfolios

| Stocks | Min Risk | Opt Risk |
|---|---|---|
| RIL | 41.18 | 41.95 |
| BPC | 12.79 | 4.27 |
| ONG | 16.10 | 3.79 |
| ATG | 2.98 | 49.68 |
| GAI | 2.69 | 0.30 |
| Portfolio Annual Return | 18.03 % | 63.94 % |
| Portfolio Annual Risk | 24.79 % | 34.68 % |

TABLE XVII  Act and Pred Returns of Oil & Gas Portfolio

| Stock | Date: Jan 1, 2021 | | | Date: Jun 1, 2021 | | | |
|---|---|---|---|---|---|---|---|
| | Amt Invstd | Act Price | No of Stock | Act Price | Act Val | Pred Price | Pred Val |
| RIL | 41950 | 1988 | 21.10 | 2169 | 45766 | 2080 | 43888 |
| BPC | 4270 | 382 | 11.18 | 471 | 5266 | 467 | 5221 |
| ONG | 3790 | 93 | 40.75 | 118 | 4809 | 114 | 4646 |
| ATG | 49680 | 377 | 131.78 | 1441 | 189895 | 1295 | 170655 |
| GAI | 310 | 124 | 2.50 | 160 | 400 | 158 | 395 |
| Total | 100000 | | | | 246136 | | 224805 |
| ROI | Actual: 146.14 %  Predicted: 124.81 % | | | | | | |

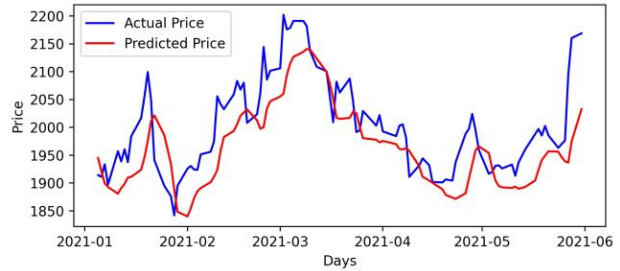

Figure 8. The act vs the pred values of Reliance Industries (RIL) stock predicted by the LSTM model for the period: Jan 1– Jun 1, 2021.

### F. Oil and Gas Sector

The five most significant stocks in this sector and their respective weights (in percent) in computing the sectoral index are Reliance Industries (RIL): 31.24, Bharat Petroleum Corporation (BPC): 11.15, Oil and Natural Gas Corporation (ONG): 10.50, Adani Total Gas (ATG): 9.39, and GAIL

India (GAI): 7.31 [2]. Tables XVI and XVII depict the results of this sector's portfolios. Fig. 8 depicts the actual and predicted prices for the leading stock, RIL.

TABLE XVIII   FMCG SECTOR PORTFOLIOS

| Stocks | Min Risk | Opt Risk |
|---|---|---|
| HUL | 0.2820 | 0.3758 |
| ITC | 0.2778 | 0.0052 |
| NST | 0.2613 | 0.1747 |
| BRT | 0.1226 | 0.0127 |
| TCP | 0.0564 | 0.4315 |
| Portfolio Annual Return | 23.75 % | 45.99 % |
| Portfolio Annual Risk | 17.86% | 22.14 % |

TABLE XIX   ACT AND PRED RETURNS OF FMCG PORTFOLIO

| Stock | Date: Jan 1, 2021 | | | Date: Jun 1, 2021 | | | |
|---|---|---|---|---|---|---|---|
| | Amt Invstd | Act Price | No of Stock | Act Price | Act Val | Pred Price | Pred Val |
| HUL | 37579 | 2388 | 15.74 | 2358 | 37115 | 2305 | 36281 |
| ITC | 525 | 214 | 2.45 | 215 | 527 | 216 | 529 |
| NST | 17470 | 18451 | 0.95 | 17759 | 16871 | 17385 | 16516 |
| BRT | 1274 | 3568 | 0.36 | 3447 | 1241 | 3442 | 1239 |
| TCP | 43152 | 602 | 71.68 | 666 | 47739 | 673 | 48241 |
| Total | 100000 | | | | 103493 | | 102806 |
| ROI | Actual: 3.49 %   Predicted: 2.81 % | | | | | | |

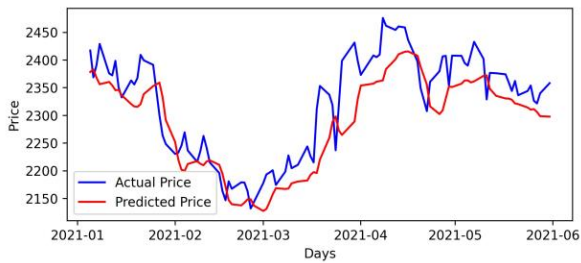

Figure 9.  The act vs the pred values of Hindustan Unilever (HUL) stock predicted by the LSTM model for the period: Jan 1– Jun 1, 2021.

TABLE XX   THE SUMMARY OF THE RESULTS

| Portfolio | Pred Return (%) | Act Return (%) |
|---|---|---|
| Auto | -0.37 | -0.51 |
| Cons. Durable | 26.80 | 27.20 |
| Healthcare | 15.34 | 14.72 |
| IT | 16.77 | 15.59 |
| Metal | 141.36 | 153.41 |
| Oil and Gas | 124.81 | 146.14 |
| FMCG | 2.81 | 3.49 |

*G. FMCG Sector*

The five most impactful stocks and their respective weights in the computation of the overall sectoral index for this sector are Hindustan Unilever (HUL): 27.59, ITC (ITC): 25.00, Nestle India (NST): 8.34, Britannia Industries (BRT): 5.70, and Tata Consumer Products (TCP): 5.57 [2]. Tables XVIII and XIX present the performances of the portfolios. Fig. 9 exhibits the actual and the predicted prices of HUL.

*H. Summary of the Results*

The results are summarized in Table XXIV which depicts the actual and the predicted returns for all seven portfolios. It is observed that while the metal has yielded the highest rate of return over the five months (i.e., Jan 1, 2021, to Jun 1, 2021), the only sector that yielded a negative return is the auto sector. Also, the LSTM model is highly accurate.

VI.   CONCLUSION

We have presented seven optimized portfolios for seven critical sectors of India based on the historical stock prices from Jan 1, 2010, to Dec 31, 2020. We also designed an LSTM model for predicting future stock prices. After a hold-out period of five months, we compute the actual and the predicted return of each portfolio and compare their values to evaluate the accuracy of the LSTM. The model is found to be highly accurate in predicting stock prices over a short horizon.